# Rethinking Security Incident Response: The Integration of Agile Principles

**George Grispos**
University of Glasgow
g.grispos.1@research.gla.ac.uk

**William Bradley Glisson**
University of South Alabama
bglisson@southalabama.edu

**Tim Storer**
University of Glasgow
Timothy.Storer@glasgow.ac.uk

## Abstract[1]

In today's globally networked environment, information security incidents can inflict staggering financial losses on organizations. Industry reports indicate that fundamental problems exist with the application of current linear plan-driven security incident response approaches being applied in many organizations. Researchers argue that traditional approaches value containment and eradication over incident learning.

While previous security incident response research focused on best practice development, linear plan-driven approaches and the technical aspects of security incident response, very little research investigates the integration of agile principles and practices into the security incident response process. This paper proposes that the integration of disciplined agile principles and practices into the security incident response process is a practical solution to strengthening an organization's security incident response posture.

*Keywords*

Information Security, Incident Response, Agile Incident Response, Agile Manifesto, Agile Principles

## 1. Introduction

Organizations are reliant on information systems in today's globally networked environments. However, industry reports indicate that security incidents continue to target these systems and are inflicting staggering financial losses on organizations (Kaspersky 2013; Ponemon Institute 2013a; PricewaterhouseCoopers 2014). The 2014 PricewaterhouseCoopers Global State of Information Security Survey (2014) claims that the rate of security incidents detected in the past twelve months increased by more than 25% and more than doubled since 2011. The Ponemon 2013 Global Cost of Data Breach survey (Ponemon Institute 2013a) and the Kaspersky 2013 Global Corporate IT Security Risks survey (Kaspersky 2013) estimate that the financial losses from security incidents and data breaches are in the millions of dollars within the past year.

In an effort to address information security attacks and data breaches, many organizations have chosen to create security incident response teams (Killcrece et al. 2003). The primary goal of a security incident response team is to minimize the effects of an incident along with managing the organization's return to an acceptable security posture (Wiik et al. 2005). Several organizations, such as the National Institute of Standards and Technology (Cichonski et al. 2012), the European Network and Information Security Agency (2010) and the International Organization for Standardization (British Standards Institution 2011), have published guidance on security incident investigation and recovery techniques. In addition to best practices, academic researchers have developed security incident processes (Mandia et al. 2003; Mitropoulos et al. 2006; Vangelos 2011; Werlinger et al. 2007). However, many of these security incident response approaches are based on a linear plan-driven model where preparation leads to the detection of an incident, followed by its containment which, in turn, allows incident response teams to eradicate, recover and then, potentially, provide feedback information into the preparation stage.

---

[1] **Please cite this paper as:** George Grispos, William Bradley Glisson and Tim Storer (2014). *Rethinking Security Incident Response: The Integration of Agile Principles*. 20th Americas Conference on Information Systems (AMCIS 2014), Savannah, Georgia.

Recent commercial deliberations argue that fundamental problems exist with the application of current approaches in real-world security incident handling context (FireEye 2013; Ponemon Institute 2013b; SANS 2013). The concerns focus on current solutions being too slow, ineffective, not providing enough insight into the causes of the incident, focusing more on containment and eradication along with undermining the value of evidence that may be required for subsequent legal action (FireEye 2013; Ponemon Institute 2013b; SANS 2013). An additional study corroborates these issues indicating that 85% of respondents stipulate that little to no prioritization of security incidents was taking place, 60% noted that there were too many manual steps to follow in their process and, just under, 55% identified that investigating incidents took too long (Ponemon Institute 2014).

As security incidents increasingly impact organizations, it is imperative that organizations have the ability to investigate, report and, ultimately, improve overall security efforts based on previous security incidents. This paper proposes the integration of disciplined agile principles and practices into the security incident response process as a viable method of strengthening an organization's security incident response proficiencies.

The structure of the paper is as follows. Section two presents industry and academia security incident response approaches along with identifying industrial incident investigations challenges. Section three identifies agile process requirements. Section four discusses how disciplined agile principles and practices can be integrated into the security incident response process. Section five draws conclusions and presents recommendations for future research.

## 2. Analysis of Security Incident Response Issues

Both practitioners and academic researchers have noted the financial impact of security breaches along with investigating ways to address increased security incidents in organizations (Glisson et al. 2006). Zafar, et al. (2012) performed a study to investigate the impact when an organization publicly announces an information security breach. The results from the study suggest that a security breach announcement not only affects the impacted organization, but can also have an effect on the wider industry as a whole (Zafar et al. 2012).

Hence, security incident response, for the purpose of this discussion, is defined as the process where a dedicated or ad-hoc incident response team detects, investigates, eradicates and recovers from a security incident (West-Brown et al. 2003; Wiik and Kossakowski 2005). Organizations have been researching and examining different solutions to formalize security incident response processes in order to implement a more proactive incident response posture (Killcrece et al. 2003). As a result, numerous security incident response approaches and best practice guidelines have been published in industry (British Standards Institution 2011; Cichonski et al. 2012; European Network and Information Security Agency 2010; Northcutt 2003; West-Brown et al. 2003) and academia (Mitropoulos et al. 2006; Vangelos 2011; Werlinger et al. 2007).

Many of these approaches are structured using a linear plan-driven approach. In these types of approaches, preparation leads to detection which is followed by containment that permits eradication and feedback into the next preparation stage. Although much of the literature has focused on the technical practices for implementing security incident response capabilities within organizations, researchers have also identified and discussed several problems with these current processes. These criticisms focused on approaches being too linear, not reflecting the concurrent lifecycle of real-world incident handling, not providing enough insight into the causes of the incident and not maximizing the benefits of digital forensic capabilities (Ahmad et al. 2012; Casey 2005; Casey 2006; Grimes 2007; Shedden et al. 2010; Shedden et al. 2011; Tan et al. 2003; Werlinger et al. 2007). The security incident response challenges identified in the literature are summarized in Table 1 – Security Incident Response Issues.

| Security Incident Response Issue | Citation(s) |
|---|---|
| The traditional linear incident response model does not support the highly efficient capability that is required to handle and manage today's incidents. | (Gonzalez 2005; Grimes 2007; Werlinger et al. 2010) |
| There is a progression flaw in linear processes, if one phase in the linear process is not completed, the entire process cycle may stop midstream. | (Grimes 2007) |
| Important steps, are often skipped because the incident response process is too focused on containment, eradication, and recovery. | (Ahmad et al. 2012; Grimes 2007; Tan et al. 2003) |
| Current approaches do not provide enough insight into the underlying causes of the incident. | (Ahmad et al. 2012; Jaatun et al. 2009; Shedden et al. 2010; Shedden et al. 2011) |
| Poor provisions for incident planning. | (Tan et al. 2003) |
| Do not maximize the benefits of digital forensic capabilities. | (Casey 2005; Casey 2006) |
| Undermine the value of forensic evidence possibly required for subsequent legal action. | (Casey 2005; Tan et al. 2003) |

**Table 1. Security Incident Response Issues**

Grimes (2007) argued that the traditional linear incident response model has become outdated and does not support the highly efficient capability that is required to handle and manage today's incidents. Grimes (2007) proclaims that although organizations have implemented the traditional linear incident model, he suspects that they do not follow it effectively. In addition, Grimes notes that there is a progression flaw in linear processes. If one phase in the linear process is not completed, the entire process cycle may stop midstream (Grimes 2007). Emphasizing the importance to incident completion and resolution, Gonzalez (2005) claims that there has been a technological shift in the way security incidents are affecting organizations and, as a result, the fast and accurate detection and resolution of incidents is a critical ability for many organizations. The main reasons for this shift is that attackers are now using automated tools to extend attacks and the unavailability of important information systems can result in significant damage to reputations, sensitive information and financial losses (Gonzalez 2005).

Werlinger, et al. (2010) focused on the socio-technical aspects of incident response and explored the security incident activities of practitioners across various industries and organizations. He did this to determine what tools were used in the incident response process and how such tools could be improved. Werlinger, et al. (2010) concluded that current incident response guidelines and tools do not appropriately support the highly collaborative nature of incident response investigations and that incident handlers often need to develop their own tools to perform specific exploratory tasks.

Ahmad, et al., (2012), Shedden, et al (2010; 2011) and Jaatun, et al (2009) focused on problems with current incident response approaches in reference to incident learning. Ahmad, et al. (2012) noted that steps, such as lessons learned, are often skipped because the incident response process is too focused on containment, eradication, and recovery. Ahmad, et al., (2012) performed a case study in an organization examining the shortcomings of incident response in practice. Their case study revealed that, although the organization in their study closely followed industry best practices, the organization's inclination was to focus on improving the technical aspects of security incidents along with maintaining business continuity. The authors highlighted the fact that the organization neglected to engage any post-incident learning activities (Ahmad et al. 2012). This is a finding that is shared by Shedden, et. al., (2011), who argued that incident response practices in organizations are highly informal and, as a result, the learning from security incidents should also be informal. Shedden, et. al.,

(2011) proposed the "Informal and Incidental Learning Model" to encourage security incident learning within organizations. Similarly, Jaatun, et. al., (2009) proposed the Incident Response Management (IRMA) method, which draws on the NIST 800-61 and ISO 27035 models, with an increased emphasis on proactive preparation and reactive learning. To achieve this, the learning phase of IRMA focuses on learning from an incident by identifying sequences of events using the Sequential Timed Events Plotting (STEP) method (Hendrick and Benner 1986). However, the model was developed to specifically handle and investigate information security incidents in the petroleum industry.

Casey (2006) argued that implementation of digital forensic practices is evolving from merely a customary role in law enforcement to a more comprehensive resolution for organization's to investigate prohibited acts. However, researchers have indicated that organizations may not be maximizing corporate forensic capabilities along with undermining the value of forensic evidence, potentially, required for subsequent legal action (Casey 2005; Nnoli et al. 2012; Tan et al. 2003). Nnoli (2012) contended that a lack of forensic readiness could result in organizations wasting effort, time and financial resources when conducting forensic investigations. The proper collection of forensic evidence, potentially, benefits organizations through faster incident resolution, legal defense support, demonstration of due diligence and verification of commercial transactions (Nnoli et al. 2012).

Casey (2005) claimed that even with a moderate amount of forensic preparation, an organization can mitigate the impact of an incident and can enable the organization to pursue legal action, if it is required. Casey (2005) adds that closer collaboration is required between security incident handlers, system administrators, and forensic examiners, so that all relevant roles understand the need to report even seemingly minor security incidents. Separately, Casey (2006) states that forensic investigations in corporate environments can be challenging because few logging systems are designed with evidentiary value in mind and that forensic specialists must apply the principles of evidence preservation creatively to each source of log data that an organization maintains.

Tan, et al. (2003) explored factors which influenced security managers to not conduct security incident investigations. These factors included a highly regulated industry which penalizes organizations for security incidents, a lack of prior planning and industrial emphasis on system recovery as opposed to performing an incident investigation (Tan et al. 2003). Furthermore, Tan, et al. (2003) reported that the organization, in their case study, did not have a clear definition for a security incident. They also noted that the organization was unaware of the benefits associated with prosecuting offenders related to security incidents or identifying how the attack took place.

While previous security incident response research focused on best practice development, linear plan-driven approaches and incident response technical aspects, minimal empirical research investigates the integration of agile principles and practices into the security incident response process.

## 3. Agile Integration

In the context of software development, Sidky and Arthur (2007) have argued that many organizations adopt agile processes to take advantage of the numerous benefits that it offers to an organization. They state that these include: reduced costs, better software quality, higher customer satisfaction, and a more responsive approach to dynamic market conditions. Agile is a term which has traditionally been reserved to classify a number of methodologies in software engineering, including Extreme Programming (XP) (Beck and Andres 2004), Scrum (Johnson 2003) and the Crystal family of agile processes (Cockburn 2004). However, agile principles have also been integrated into security solutions like the Web Engineering Security Methodology (Glisson 2008).

The Agile Manifesto and its twelve underlying principles characterize agile processes (Beck et al. 2001). This Manifesto and agile principles are the essential characteristics that must be reflected in a process before it is considered agile (Sidky et al. 2007). Security incident response issues identified in the literature suggests that security incident response processes are a prime candidate for the integration of agile values and principles into the process:

- **Individuals and interactions are valued over processes and tools** (Beck et al. 2001). The Agile Manifesto suggests that agile processes are very reliant on individuals and the interaction between them (Beck et al. 2001). Werlinger, et al (2010) and Casey (2005) have argued that individuals are one of the key factors in the success or failure of security incident investigations. Thus, the agile route with its focus on people lends itself strongly to the security incident response process.
- **Working software over comprehensive documentation** (Beck et al. 2001). Agile processes, for software development, change the focus from developing volumes of documentation to focusing on the delivery of working software systems (Beck et al. 2001). Similarly, security researchers (Ahmad et al. 2012; Grimes 2007; Tan et al. 2003) have suggested that organizations are currently more focused on containment, eradication, and recovery than security incident learning to prevent reoccurrence. Although containment, eradication, and recovery are important deliverables for a security incident investigation, this approach should be expanded to include the opportunity for an organization to learn from an incident and mitigate chances for reoccurrence.
- **Customer collaboration over contract negotiation** (Beck et al. 2001). Agile processes value substantial interaction between customers, developers, and project stakeholders to ensure that the product being developed satisfies the customer's business needs (Beck et al. 2001). Similar calls by Casey (2005) and Werlinger (2010) for cooperation between various security, technology and legal roles for forensic investigations within an organization, supports the idea that a security incident response process would benefit from forming partnerships and increasing collaborative efforts between incident response teams and other roles within an organization.
- **Responding to change over following a plan** (Beck et al. 2001). The ability to respond quickly to change is viewed as an asset in an agile process (Beck et al. 2001). The integration of agile principles and practices could assist in providing a security incident response team with the ability to respond quickly to change in contrast to following a linear plan-driven approach to incident handling. Many traditional approaches require the identification of potentially compromised assets at the beginning of the incident lifecycle and restricts their return to the production environment to the end of the investigation (Mitropoulos et al. 2006). The adoption of an iterative and incremental approach to incident handling could escalate restoration of assets to the production environment while continuing the investigation lifecycle iterations.

## 4. Agile Security Incident Response Improvements

Disciplined agile principles and practices could provide a solution to the security incident response challenges identified in the literature through iterative and incremental incident handling, reducing uncertainty, and through continuous attention to technical excellence.

### 4.1 Iterative and Incremental Incident Handling

Rubin (2012) argues that one problem with plan-driven processes is that it requires that individuals get things right up-front. Similarly, many of the current security incident response processes (British Standards Institution 2011; Cichonski et al. 2012; European Network and Information Security Agency 2010; Mitropoulos et al. 2006), have been designed as linear plan-driven processes that require up-front identification of assets to be included in the subsequent phases of the incident response lifecycle. If further assets are identified, the process needs to begin again from the identification phase of the process. As a result, assets included in the investigation are kept offline for majority of the incident lifecycle and their return to the production environment is only orchestrated in the recovery phase of the lifecycle. For many traditional security incident response processes, this recovery phase is usually at the end of the linear process.

An iterative and incremental approach towards security incident handling provides an incident response team with an opportunity to satisfy asset stakeholders through the early and continuous restoration of information assets. This can be achieved through a continual process of prioritization and reprioritization of assets. This provides a security incident response team with the opportunity to iteratively reevaluate an asset's required participation in an investigation. If the asset is no longer required, it can be returned to the production environment in the subsequent iteration. Figure 1 presents a potential implementation of this idea.

The concept is that a list of potential assets will be provided to the Security Incident Response Team (SIRT) and relevant stakeholders. Stakeholders in this context undertake a similar role to customers in Extreme Programming and product owners in Scrum in that they define the investigation and set its goals. Ultimately, it is the stakeholders who will decide the success or failure of an incident investigation. The first step (1) is for the SIRT and stakeholders to analyze and evaluate known incident information. The second step (2) is based on the analysis preformed in step one. The stakeholders prioritize the importance of each asset to the overall investigation. The process of prioritization allows the SIRT and stakeholders to reference an existing asset registry in order to identify known and, potentially, unknown assets that have been compromised and may need to be investigated. The SIRT then estimates the time required for asset resolution on one or more assets in step three (3). Step four (4) allocates action responsibilities to relevant individuals.

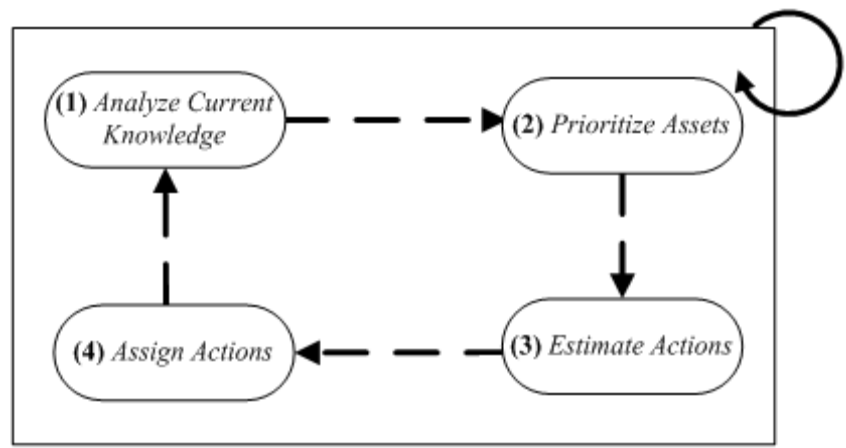

Figure 1. Iterative and Incremental Incident Response

When the assigned actions in step four are complete, the SIRT evaluates the current incident knowledge based on the investigation information from the assets in the previous iteration. This provides the team with an opportunity to reprioritize assets and assess their current value to the investigation. If the incident response team is satisfied that the asset is no longer important to the investigation, it can be returned to the production environment. Likewise, in step four, assigned actions could identify additional assets for investigation. These assets are added to the investigation and the current list of prioritized assets can be reprioritized to reflect this change. This iterative and incremental approach to incident handling means that a security incident response team embraces changing requirements through the addition and removal of assets at any point during the incident lifecycle.

**4.2 Reducing Uncertainty**

Tan, et al., (2003) reported that their studied organization did not have a clear definition for a security incident and, as a result, the organization's incident response team was not sure when to perform a full security investigation. An analysis of the industry's best practices indicates that a consensus on the definition of a 'security incident' has not emerged and is perceptible in the following definitions:

- "Events which can cause a breach of security or a loss of integrity that would have an impact on the operation of electronic telecommunication networks and services" (European Network and Information Security Agency 2010).
- "An adverse event in an information system, and/or network, or the threat of the occurrence of such an event" (Northcutt 2003).

- "A violation or imminent threat of violation of computer security policies, acceptable use policies, or standard security practices" (Cichonski et al. 2012).
- "Any real or suspected adverse event in relation to the security of computer systems or computer networks" (West-Brown et al. 2003).

Vague comprehension of a security incident translates into organizations consuming valuable resources on security incidents that may not be considered incidents within distinct landscapes. One of the goals of an agile process is that it aims to simultaneously reduce all forms of uncertainties (Rubin 2012). One resolution to this issue is to treat all reported security occurrences as 'events' until the last reasonable moment and then to declare them based on a clear definition classification as a 'security incident'. That is, an organization delays committing resources to an in-depth investigation until it is satisfied that it is, indeed, an incident.

To achieve this, individual organizations need to define security events and incidents within their organization along with escalation procedures. Clear definitions remove classification vagueness for the SIRT and refines investigation resource allocation. Implementing an iterative and incremental approach to event/incident investigations, along with additional analysis tools, like cost-benefit analysis, allows an incident response team to dynamically re-evaluate classifications and to allocate resources as required.

**4.3 Continuous Attention to Technical Excellence**

Security incidents are undesirable situations; nevertheless, they present an opportunity to learn about the security risks and vulnerabilities that can exist in both technical and socio-technical systems (Line et al. 2009). However, researchers claim that organizations do not pay enough attention to incident learning (Ahmad et al. 2012; Shedden et al. 2010; Shedden et al. 2011). They go on to claim that organizations are more concerned with eradication and recovery (Ahmad et al. 2012; Shedden et al. 2010; Shedden et al. 2011).

A solution to this issue is to integrate security incident learning throughout the incident lifecycle. Agile approaches emphasize continuous attention to technical excellence, as this enhances and maintains agility (Rubin 2012). The integration of data capture designed to identify event/incident sequences, root cause analysis techniques, barrier analysis, automated forensic data capture solutions, along with centralized data repositories, can increase a SIRT's agility and encourage continuous learning throughout the incident life cycle.

A root cause analysis can help establish why a security incident has occurred and a barrier analysis can be used to analyze the vulnerabilities associated with the incident and the barriers that should have been in place to prevent it (U.S. Department of Energy 2012). This form of analysis could be used to identify barriers that were in place and how they performed; barriers that were in place but not used; barriers that were not in place but were required; and the barrier(s) that, if present or strengthened, would prevent the same or similar accidents from occurring in the future (Sklet 2004).

Digital forensics solutions extract data of particular interest from an asset. Data which can typically be recovered using digital forensic techniques includes deleted, damaged or hidden files; metadata about files including file modification, access, and creation times; as well as logs and emails (Kent et al. 2006). However, the aim of an investigation should not be restricted to uncovering evidence of malicious activity, but to also examine the technical and socio-technical issues that have contributed to the incident. The tools and practices being discussed are available in the accident and safety industries to examine the technical and socio-technical issues around an incident (Johnson 2003; Sklet 2004). The use of these practices can be transferred to security incident investigations. The integration of these tools and methods into security would increase the amount of data potentially being captured and enhance the technical excellence of security incident learning. This will hopefully translate into providing a SIRT with an opportunity to produce richer security lessons from investigations.

## 5. Future Work and Conclusions

Potential deficiencies in technical and socio-technical security controls are forcing organizations to implement security incident response measures to detect, respond to, and recover from incidents. Several incident response approaches have been proposed, however, recent industrial surveys and academic research argue that deficiencies exist in the current approach towards security incident handling. A possible solution to these issues is the integration of disciplined agile principles and practices into the security incident response process. This paper highlights the integration of agile principles and practices such as iterative and incremental incident handling, reducing uncertainty, and continuous attention to technical excellence could enhance real-world security incident response efficiencies and effectiveness.

Future research will include a closer examination of the business case for the integration of disciplined agile principles and practices into the security incident response process through the implementation of exploratory organizational case studies, surveys and retrospective assessments. The case study will provide information on the incident response landscape within an organization to help understand real-world interaction and management of security incidents. In addition, organizational surveys will be used to develop a deeper understanding of the security incident response challenge facing a SIRT. The retrospective assessments will be used to acquire an in-depth awareness of how the team collaborates with the rest of the organization along with encouraging the SIRT to consider and identify areas for improvement in the, overall, agile SIRT process. The results from relevant literature, case studies, surveys and retrospective assessments will then be used to identify criteria along with providing the foundation for the development of an Agile Incident Response (AIR) methodology. Ultimately, the high-level goal is the deployment and continued refinement of the AIR methodology in multiple industries.

## 6. Acknowledgements

This work was supported by the A.G. Leventis Foundation. Any opinions, findings, conclusions or recommendations expressed in this paper are those of the authors and do not reflect the views of the A.G. Leventis Foundation.